\documentclass[preprint,prc]{revtex4}
\usepackage{epsf}
\usepackage{amsmath,amsthm,amssymb}
\usepackage{color}
\usepackage{dcolumn}
\usepackage{bm}
\usepackage{graphicx,epsfig}

\graphicspath{{fig/}}
\everymath{\displaystyle}

\begin{document}           

\title{Corrections for a constant radial magnetic field in the muon $\bm g$--2 and electric-dipole-moment experiments in storage rings}

\author{Alexander J. Silenko} 
\affiliation{Research Institute for Nuclear Problems, Belarusian State
University, Minsk 220030, Belarus} \affiliation{
Bogoliubov Laboratory of Theoretical Physics, Joint Institute for Nuclear Research, Dubna 141980, Russia}

\date{\today}

\begin {abstract}
We calculate the corrections for constant radial magnetic field in muon \emph{g}--2 and electric-dipole-moment experiments in storage rings. While the correction is negligible for the current generation of \emph{g}--2 experiments, it affects the upcoming muon electric-dipole-moment experiment at Fermilab.
\end{abstract}

\maketitle

\section{Introduction}

Muon \emph{g}--2 and electric-dipole-moment (EDM) experiments in storage rings are of great importance for contemporary physics. The well-known discrepancy between theoretical and experimental values of the anomalous magnetic moment (AMM) of muon \cite{PRDfinal,gmtFermi} may be a window to new physics. New physics can result in a higher value of the muon EDM than that predicted by the Standard Model. 

In such precise experiments it is important to fully account for all significant corrections affecting the measurement. This is the reason for conducting the present research which determines corrections for a constant radial magnetic field in high-precision \emph{g}--2 and EDM experiments with polarized muon beams in storage rings. Increasing of accuracy of the planned experiments as compared to the accomplished ones makes this research to be necessary.


We use the cylindrical coordinate system and the system of units $\hbar=1,~c=1$. We include $\hbar$ and
$c$ into some equations when this inclusion clarifies the problem.

\section{General equations of spin motion in the cylindrical and the Frenet-Serret coordinate systems}\label{CylFreSe}

In this section, we shortly report the results of comparison of the cylindrical and the Frenet-Serret coordinate systems obtained in Ref. \cite{JINRLettCylr}. While these coordinate systems are equivalent, the related equations of the spin motion differ. The axes of the Frenet-Serret coordinate system are motional and are defined relative to the particle velocity direction. In this system, the equation of spin motion defines a motion of the spin (pseudo)vector with
respect to the momentum vector, i.e., the change in a relative orientation of these vectors. The Frenet-Serret coordinate system is commonly used in accelerator theory. For a particle with electric and magnetic dipole moments, the classical equation of the spin motion is given by (see Refs. \cite{GBMT,PhysScr} and references therein)
\begin{eqnarray}
\bm\Omega^{(FS)}=-\frac{e}{m}\left[a\bm B-\frac{a\gamma}{\gamma+1}\bm\beta(\bm\beta\cdot\bm B)+\left(\frac{1}{\gamma^2-1}-a\right)\left(\bm\beta\times\bm
E\right)\right.\nonumber\\
+\left.\frac{\eta}{2}\left({\bm E}-\frac{\gamma}{\gamma+1}\bm\beta(\bm\beta\cdot\bm E)+{\bm\beta}\times {\bm B}\right)\right],
\label{Nelsonh}\end{eqnarray} where $a=(g-2)/2, ~\bm\beta=\bm v/c, ~g=2mc\mu/(e\hbar s), ~\eta=2mcd/(e\hbar s)$. Here $s$ is the spin quantum number and the $\eta$ factor for the EDM $d$ is similar to the $g$ one for the magnetic dipole moment.

When the cylindrical coordinate system is used, the spin motion equation is defined allowing for a motion of the horizontal axes of the cylindrical coordinate system which follows the azimuthal motion of a given particle. The vertical axes of the cylindrical and the Cartesian coordinate systems always coincide. The angular velocities of the spin motion in these coordinate systems differ on $\dot{\phi}\bm e_z$, where $\dot{\phi}$ is the angular velocity of a change of the particle azimuth in the storage ring \cite{RPJSTAB}. As a result, the spin motion equation in the cylindrical coordinate system takes the form \cite{JINRLettCylr,RPJSTAB}
\begin{eqnarray}
\bm\Omega^{(cyl)}=-\frac{e}{m}\left\{a\bm B-
\frac{a\gamma}{\gamma+1}\bm\beta(\bm\beta\cdot\bm B)\right.\nonumber\\
+\left(\frac{1}{\gamma^2-1}-a\right)\left(\bm\beta\times\bm
E\right)+\frac{1}{\gamma}\left[\bm B_\|
-\frac{1}{\beta^2}\left(\bm\beta\times\bm
E\right)_\|\right]\nonumber\\ \left.+ \frac{\eta}{2}\left(\bm
E-\frac{\gamma}{\gamma+1}\bm\beta(\bm\beta\cdot\bm
E)+\bm\beta\times\bm B\right)\right\}. 
\label{eq7}\end{eqnarray} The sign $\|$ indicates the horizontal
projection for any vector.
This equation describes the spin motion relative to immobile detectors placed around the storage ring. Equation (\ref{eq7}), contrary to Eq. (\ref{Nelsonh}), does not depend on a vertical motion of the beam.
The compactness of Eq. (\ref{Nelsonh}) relies on the fact that the Frenet-Serret coordinate axes move relative to the immobile detectors. As a result, Eq. (\ref{Nelsonh}) needs to be used with care. In particular, this equation may create the illusion that, on condition that $\bm\beta\cdot\bm B=0$, the impact of vertical and radial fields on the spin is identical.
For leptons (electron, muon) the difference between $\bm\Omega^{(FS)}$ and $\bm\Omega^{(cyl)}$ can be rather significant. To determine an \emph{observable} effect, the motion of the tangential and normal axes
of the Frenet-Serret coordinate system should be added to the spin motion in
this coordinate system. Once this factor is taken
into account, the cylindrical and the Frenet-Serret coordinate systems give an equivalent description of the spin motion \cite{JINRLettCylr}.

The additional term in Eq. (\ref{eq7}) which is proportional to $1/\gamma$ can increase the importance of the horizontal components of $\bm B$ and $\bm\beta\times\bm E$ as compared to Eq. (\ref{Nelsonh}). In particular, the spin quantization axis (also called the stable spin axis) in a purely magnetic storage ring does not coincide with the direction of the total magnetic field. In relation to the longitudinal magnetic field, this property has been considered in Ref. \cite{JETPazm}.

It should be added that the quantum-mechanical description of the spin motion for spin-1/2 \cite{RPJ} and spin-1 \cite{PRDspin} particles fully agrees with the classical description.

\section{Manifestations of the constant radial magnetic field in muon $\bm g$--2 and electric-dipole-moment experiments}\label{Manifestations}

The correction for the constant radial magnetic field is nonzero in both the \emph{g}--2 and electric-dipole-moment experiments with polarized muons in storage rings. However, the importance of the systematic effects caused by this field significantly differs in the two cases.
For a calculation of this correction, it is convenient to use Eq. (\ref{eq7}). Since the \emph{average} Lorentz force acting on a particle is zero, the average values of the vertical electric field and the radial magnetic one satisfy the condition
\begin{eqnarray}
E_z=-(\bm\beta\times\bm B)_z.
\label{eqLf}\end{eqnarray} When electric focusing is used, the angular velocity of the spin precession is given by
\begin{eqnarray}
\bm\Omega^{(cyl)}=-\frac{e}{m}\left\{aB_z\bm e_z+\frac{g}{2\gamma^2}B_\rho\bm e_\rho
+ \frac{\eta}{2}\left(\bm\beta\times\bm B\right)_\|\right\}.
\label{eqcel}\end{eqnarray}

It is well-known that the fields satisfying Eq. (\ref{eqLf}) turn the spin. The vertical electric field conditions one of main systematic errors in EDM experiments in storage rings carried out by the frozen spin method \cite{FJM}. The condition (\ref{eqLf}) is satisfied for Wien filters with static \cite{TA} and oscillating \cite{rfWienFilter,rfWienFilterNIMA,PhysRevSTAB2017} fields which are frequently used as spin rotators. In particular, the expression for the radial component of $\bm\Omega$ (halved owing the field oscillations) is presented in Ref. \cite{rfWienFilter}.

In the muon \emph{g}--2 and EDM experiments described in Refs. \cite{PRDfinal,MuEDM08}, $\gamma^2=1+a^{-1}$ and $g/(2\gamma^2)=a$. As a result, Eq. (\ref{eqcel}) takes the form
\begin{eqnarray}
\bm\Omega^{(cyl)}=-\frac{e}{m}\left\{a(B_z\bm e_z+B_\rho\bm e_\rho)
+ \frac{\eta}{2}\left(\bm\beta\times\bm B\right)_\|\right\}.
\label{eqfin}\end{eqnarray}

The angular velocity of the spin rotation caused by the magnetic moment is collinear to the vector of the total magnetic field $\bm B$. However, the vertical and radial components of $\bm B$ enter Eq. (\ref{eqcel}) with different factors and the spin motion defined by Eq. (\ref{eqfin}) is influenced not only by the radial magnetic field but also by the vertical electric one.

Equation (\ref{eqfin}) can be presented as follows:
\begin{eqnarray}
\bm\Omega^{(cyl)}=\omega_0\bm e_z+O\bm e_\rho, \qquad \omega_0=-\frac{e}{m}aB_z,
\qquad O=\omega_0\left(\frac{B_\rho}{B_z}
+ \frac{\eta\beta}{2a}\right),
\label{eqfes}\end{eqnarray} where the quantities $\omega_0, \,O$, and $\beta$ ($\bm\beta=\beta\bm e_\phi$) can be positive and negative.

Evidently, the constant radial magnetic field conditions the following correction to the \emph{g}--2 frequency:
\begin{eqnarray}
\Delta\omega=-\omega_0\frac{B_\rho^2}{2B_z^2}.
\label{eqgmt}\end{eqnarray}

In the \emph{g}--2 experiments \cite{PRDfinal,gmtFermi}, this correction can be neglected. Measurements of the constant radial magnetic field in the BNL muon \emph{g}--2 experiment have shown that $|B_\rho|\approx 2.9\times10^{-5}$T. Since $B_z= 1.45$T,
$|B_\rho/B_z|\approx 2\times10^{-5}$ \cite{BNLrmfm}. In this case, the correction to the \emph{g}--2 frequency is of the order of $10^{-10}$ and can be neglected.

However, the constant radial magnetic field can be more important for the muon EDM experiments. The search for the muon EDM has been performed in the framework of the past BNL experiment \cite{BNLrmfm}. The initial beam polarization was longitudinal and the vertical spin component
was detected. This experiment has established the upper bound on the muon EDM, $|d_\mu|< 1.8 \times10^{-19}~e\cdot$cm  \cite{MuEDM08}.

In the considered case, squares of small parameters can be neglected. The spin evolution is defined by the general equation (15) in Ref. \cite{EPJC2017} which takes the form
\begin{equation}
\begin{array}{c}
P_\rho(t)=\cos{(\omega_0 t+\psi)},\qquad
P_\phi(t)=\sin{(\omega_0 t+\psi)},\qquad
P_z(t)=-\frac{O}{\omega_0}\cos{(\omega_0 t+\psi)},
\end{array}\label{polvmof} \end{equation}
where $\psi$ is equal to $\pi/2$ and $-\pi/2$ for counterclockwise and clockwise muon beams, respectively, and $\bm P$ is the polarization vector.
In this case, a nonzero vertical component of the polarization vector is conditioned by the terms \emph{linear} in $B_\rho/B_z$ and $\eta$. The related correction to the \emph{g}--2 frequency is \emph{quadratic or bilinear} in $B_\rho/B_z$ and $\eta$.

Equations (\ref{eqfes}) and (\ref{polvmof}) demonstrate that the constant radial magnetic field can mimic an EDM. The constant radial magnetic field measured in Ref. \cite{BNLrmfm} corresponds to the muon EDM of $2\times10^{-21}~e\cdot$cm. This value is small as compared with the current bound on the muon EDM established in Ref. \cite{MuEDM08}. However, the sensitivity to the muon EDM will be increased by two orders of magnitude in the future experiment at Fermilab \cite{FermilabEDM}. In this case, the correction for the constant radial magnetic field needs to be considered. We can add that a muon EDM experiment based on the frozen spin method \cite{FJM} is sensitive to the muon EDM of the order of $10^{-24}~e\cdot$cm.


\section{Summary}\label{summary}

In the present paper, we have calculated the correction for the constant radial magnetic field in the \emph{g}--2 and electric-dipole-moment experiments with muons in storage rings. 
While the correction can be neglected in the \emph{g}--2 experiments \cite{PRDfinal,gmtFermi}, it is important for the forthcoming muon EDM experiment at Fermilab.


The planned \emph{g}--2/EDM experiment at J-PARC will be fulfilled in a purely magnetic ring \cite{JPARC,JPARCOtani}. In this case, the average constant radial magnetic field is equal to zero and the correction for this field vanishes.

\section*{Acknowledgements}

The author is grateful to N. Nikolaev for useful
comments and discussions. The author acknowledges the support by the Belarusian Republican Foundation for Fundamental Research
(Grant No. $\Phi$16D-004) and
by the Heisenberg-Landau program of the German Ministry for
Science and Technology (Bundesministerium f\"{u}r Bildung und
Forschung).

\end{document}